\documentclass[12pt]{article}
\usepackage{amsmath,amsfonts,amsthm,amssymb,amscd,latexsym, bm, bbm}
\usepackage{hyperref}
\usepackage[usenames,dvipsnames]{color}
\usepackage{authblk}
\hypersetup{colorlinks,
	citecolor=Blue,
	linkcolor=Blue,
	urlcolor=Blue}

\usepackage[T2A]{fontenc}
\usepackage{graphicx}
\usepackage{mathrsfs}
\numberwithin{equation}{section}

{\theoremstyle {definition} \newtheorem {definition} {Definition} [section] }
{\theoremstyle {plain}  \newtheorem {theorem} [definition] {Theorem}}
{\theoremstyle {plain}  \newtheorem {corollary} [definition]{Corollary}}
{\theoremstyle {plain}  \newtheorem {lemma} [definition]{Lemma}}
{\theoremstyle {plain}  \newtheorem {proposition} [definition]{Proposition}}
{\theoremstyle {plain} }
{\theoremstyle {definition} }

\newcommand{\R}{ \mathbb{R} }

\newcommand{\supp}{\operatorname{supp}}

\newcommand{\Del}{\Delta}

\newcommand{\om}{ \omega }

\newcommand{\Om}{ \Omega }

\newcommand{\ga}{\gamma }

\newcommand{\s}{ \sigma }

\newcommand{\ka}{ \kappa }

\renewcommand{\phi}{ \varphi }

\newcommand{\eps}{\varepsilon}

\newcommand{\de}{ \delta }
\newcommand{\al}{ \alpha }

\newcommand{\Id}{\operatorname{Id}}

\newcommand{\be}{\begin{equation}}
\newcommand{\ee}{\end{equation}}
\newcommand{\ben}{\begin{equation*}}
\newcommand{\een}{\end{equation*}}

\newcommand{\ds}{\displaystyle}

\newcommand{\lan}{ \langle }

\newcommand{\blan}{ \big\langle }
\newcommand{\ran}{ \rangle}

\newcommand{\bran}{ \big\rangle}

\newcommand{\p}{ \partial}

\newcommand{\lbl}{\label}
\newcommand{\non}{\nonumber}
\newcommand{\qu}{\quad}
\newcommand{\qmb}{\quad\mbox}
\newcommand{\qnd}{\qmb{and}\qu}

\newcommand{\Ga}{\Gamma}
\newcommand{\sm}{\setminus}
\newcommand{\Ss}{\Sigma_0}

\title{Asymptotic expansions for a class of singular integrals emerging in non-linear wave systems}
\author{Andrey V. Dymov\thanks{dymov@mi-ras.ru}}
\affil{Steklov Mathematical Institute of Russian Academy of Sciences,\\ Moscow, Russia}
\date{}

\begin{document}

	\maketitle	

	\begin{abstract}
	We find asymptotical expansions as $\nu \to 0$ for integrals of the form \\ $\int_{\R^d} F(x) / \big(\om(x)^2 + \nu^2\big)\, dx$,
	 where sufficiently smooth functions $F$ and  $\om$ satisfy natural assumptions for their behaviour at infinity and all critical points of the function $\om$ from the set $\{\om(x) = 0\}$ are non-degenerate. 
	 	These asymptotics play a crucial role when analysing stochastic models for non-linear waves systems. 
	Our result generalizes that of \big[S. Kuksin, Russ. J.  Math. Phys.'2017\big] 
	where a similar asymptotics was found in a particular case when $\om$ is a non-degenerate quadratic form of the signature $(d/2,d/2)$ with even $d$.

\end{abstract}

\section{Introduction}

\subsection{Set up and result}

We study asymptotic behaviour of integrals 
\be\lbl{set_1}
\int_{\R^d} \frac{F(x)}{\om(x)^2 + \nu^2\Ga(x)^2}\, dx
\ee
as $\nu\to 0$, where $dx = dx_1\cdots dx_d$, $d\geq 2$ and $\Ga$, $F$, $\om$ are sufficiently smooth real-valued functions with behaviour at infinity satisfying natural assumptions formulated below, and the function $\Ga$ is strictly positive. We assume that 
on the set $\Sigma\cap \supp F$, where
\be\lbl{Sigma}
\Sigma = \{x\in\R^d:\, \om(x) = 0\},
\ee
the function $\om$ has only non-degenerate critical points 
and the number of these critical points is finite.

Integrals \eqref{set_1} arise in physical and mathematical works  on the wave turbulence theory. Their singular limits $\nu\to 0$ describe behaviour of certain physical characteristics, studying of which is the objective of the theory, see Section~\ref{s:mot} for a more detailed discussion. In physical works integrals \eqref{set_1} usually appear implicitly and become visible when performing rigorous analysis of the heuristic constructions employed in these works.

Dividing the numerator and denominator by $\Ga^2$  we see that it suffices to study the case $\Ga(x) \equiv 1$, i.e. integrals of the form
\be\non
I_\nu = \int_{\R^d} \frac{F(x)}{\om(x)^2 + \nu^2}\, dx.
\ee
Moreover, without loss of generality we assume that $\om$ has at most one critical point  on the set $\Sigma\cap\supp F$, where $\Sigma$ is defined in \eqref{Sigma}, and this point is $x=0$.

Our main result is as follows.
Denote $\Sigma_0 = \Sigma\sm \{0\}$ if $x=0$ is a critical point for $\om$ and $\Sigma_0 = \Sigma$ otherwise. The set $\Ss$ is a differentiable manifold of dimension $d-1$.
Denote by $d_\Sigma x$ the volume element on $\Sigma_0$ induced from $\R^d$ provided with the standard Euclidean structure and consider the integral
\be\lbl{asymp}
I_0 = \pi\int_{\Sigma_0} \frac{F(x)}{\big|\nabla \om(x)\big|}\, d_\Sigma x,
\ee
where $\big|\nabla \om\big|$ denotes the Euclidean norm of the gradient of the function $\om$. 
Corollary~\ref{c:integration} implies that the integral $I_0$ converges under assumptions A1-A4 formulated below.

Set for $r>0$ and $a,b\in\R$
\be\lbl{chi}
\chi_{a,b}(r) = \left\{
\begin{array}{cl}
	1, &\qmb{if}\qu  a \ne b, \\
	|\ln r|, &\qmb{if}\qu a = b. 
\end{array}
\right.
\ee

\begin{theorem}\lbl{t:main}
(1) Assume that $x = 0$ is a unique critical point of the function $\om$ on the set $\Sigma\cap\supp F$, this critical point is non-degenerate and 	$d\geq 4$.
	Then, under assumptions A1-A5 formulated below, 
	\be\lbl{main}
	\big| I_\nu - \nu^{-1} I_0 \big| \leq C\chi_{d,4}(\nu),
	\ee
	where the constant $C$ does not depend on $0< \nu \leq 1/2$.

	(2) Assume that the function $\om$ does not have critical points on the set $\Sigma\cap\supp F$ and $d\geq 2$. Then, under the assumptions A1-A5, $	\big| I_\nu - \nu^{-1} I_0 \big| \leq C$.
\end{theorem}

 We prove only item (1) of the theorem since the proof of item (2) can be obtained by simplification of the proof of item (1). 
We impose the restriction $d\ge 4$ since some integrals involved into analysis of $I_\nu$ strongly diverge at the critical point $x=0$ (see e.g.~ \eqref{ap_151}).

In case when $d=2n$, $n\geq 2$, and $\om$ is a non-degenerate quadratic form with index $(n,n)$ analogue of Theorem~\ref{t:main} for the integral \eqref{set_1} was proved in \cite{K}. 
Our argument follows a scheme suggested in this work, but we meet additional difficulties. In \cite{K} it is used that the set $\Sigma$ is a cone but in our case its geometry can be much more complicated. In particular, this results in the lack of explicit formulas and constructions as those employed in~\cite{K}.

In \cite[Section 6]{K} it is shown that the optimal upper bound for $I_\nu$ and related integrals may be obtained by the stationary phase method, and the proof of this bound is short. However, the asymptotic for $I_\nu$ is inconsistent with this type of methods, as it is explained in \cite[Section 3]{DNT}, where a class of fast oscillating integrals was studied by the abstract stationary phase method. 
This can be seen by observing that the main term \eqref{asymp} of the asymptotic  depends not only on values of functions $\om$, $F$ and their derivatives at the critical point $x=0,$ but on their restrictions to the whole manifold $\Sigma_0$.

\smallskip

In the next subsection we formulate assumptions, imposed on the functions $F$ and $\om$, and in Section~\ref{s:example} we consider an example when the function $\om$ is a quadratic polynomial, including the case studied in \cite{K}. In Section~\ref{s:mot} we plug our results into the context of the wave turbulence theory
thus explaining our motivation for this research. 

The rest of the paper is devoted to the proof of Theorem~\ref{t:main}.
Namely, in Section~\ref{s:neigh} we study a certain neighbourhood $U_\Theta(\Sigma_0)$ of the manifold $\Sigma_0$, introduce there appropriate coordinates, write the volume element in these coordinates and then study behaviour of various integrals over $\Sigma_0$. In Section~\ref{s:the_integral} we write the integral $I_\nu$ as a sum of three, the first one is taken over a small ball around zero, the second ~--- over the complement to the neighbourhood $U_\Theta(\Sigma_0)$, and the third one ~--- over $U_\Theta(\Sigma_0)$. We show that the first two integrals are negligible when $\nu\to 0$ and, using results of Section~\ref{s:neigh}, prove that behaviour of the third integral is governed by the desired asymptotics. 

\subsection{Assumptions} 

Below for $x\in\R^n$ we use the notation
\be\non
\lan x \ran := \max\big(1, |x|\big),
\ee
where $|x|$ denotes the Euclidean norm of $x$.

{\bf A1.} {\it The function $F$ is $C^2$-smooth and for a constant $M_F\in\R$ satisfies }
$$|\p^\al F(x)| \le C \lan x\ran^{-M_F - |\al|_1} \qmb{for any}\qu 0\leq |\al|_1\leq 2 \qmb{and any}\qu x\in\R^d,
$$ 
{\it where $\al = (\al_1,\dots, \al_d)$, $\al_j\geq 0,$ is a multi-index and $|\al|_1: = \al_1+\dots+\al_d$.}

\smallskip

{\bf A2.} {\it The function $\om$ is $C^4$-smooth and for a constant $M_\om\in\R$ satisfies }
$$|\p^\al \om(x)| \le C \lan x\ran^{M_\om - |\al|_1}
\qmb{for any}\qu 0\leq |\al|_1\leq 4 \qmb{and any}\qu x\in\operatorname{supp} F.
$$

{\bf A3.} {\it There is a constant $m_\om\in\R$ such that }
\be\lbl{A3_}
|\nabla\om(x)|\geq C|x|^{m_\om} \qmb{for any}\qu x\in\Sigma\cap \operatorname{supp} F \qmb{satisfying} \qu |x|\geq 1.
\ee

 If  $x=0$ is a non-degenerate critical point of the function $\om$, we clearly have
\be\lbl{N_xi-lower}
|\nabla\om(x)|\ge C|x| \qmb{for} \qu |x|\leq 1.
\ee

{\bf A4.} {\it The constants $M_F, M_\om$ and $m_\om$ from assumptions A1-A3 satisfy}
$$
M_F > \max( M_{cr}, 2M_{cr}) + d \qmb{where}\qu M_{cr}:= M_\om - 2m_\om -2 .
$$
Clearly, by assumptions A2 and A3 we also have
\be\lbl{M-m}
m_\om \leq M_\om - 1,
\ee
if the set $\Sigma\cap \operatorname{supp} F$ is unbounded. Otherwise the parameter $m_\om$ can be taken arbitrary, and we choose it in such a way that \eqref{M-m} is satisfied. 

\smallskip

Let $D_\kappa$, $\ka>0$, denotes the set of points $|x|\geq 1$ "well separated" from the set $\Sigma_0$:
\be\non 
D_\ka = \big\{x\in B_1^c:\, |x-\Sigma_0|\ge \ka|x|^{m_\om + 2 - M_\om} \big\},
\ee
where $B_1$ stands for the closed unit ball centred at zero and $B_1^c$\, stands for its complement, while $|x-\Sigma_0|$ denotes the Euclidean distance from $x$ to $\Sigma_0$.

{\bf A5.} {\it For any $\ka>0$ the integral
$
{\ds\int_{D_\ka} \frac{|F(x)|}{\om(x)^2} \,dx}
$
converges.}

Assumption A5 is rather implicit, so we give a sufficient condition for it.
\begin{lemma}\lbl{l:A5}
	Assume that assumptions A1-A3 are satisfied with $m_\om = M_\om - 1$ and $M_F > d-2M_\om$, and that inequality \eqref{A3_} holds for any $x\in B^c_1$ (not only for $x\in \Sigma\cap\supp F$).  Then assumption A5 is fulfilled.
\end{lemma}
{\it Proof.}
We first show that $|\om(x)| \ge C(\ka)|x|^{M_\om}$
for any $x\in D_\kappa$. We argue by contradiction assuming that 
for any $\eps > 0$ there is $x_\eps\in D_\kappa$ such that $|\om(x_\eps)| \leq \eps|x_\eps|^{M_\om}$. 
We claim that in this case there is $|\ga|<C\eps |x_\eps|^{1-m_\om}$ such that $y_\ga = x_\eps + \ga\nabla\om(x_\eps)$ satisfies $y_\ga \in \Sigma$, if $\eps$ is sufficiently small. 
Indeed, denoting $\Om = \operatorname{Hess}\om$, by the Taylor formula we find
\be\non
\om(y_\ga) = \om(x_\eps) + \ga |\nabla\om(x_\eps)|^2 + 
\frac{\ga^2}{2} \blan \Om(x_\eps + \hat\ga\nabla\om(x_\eps)) \nabla\om(x_\eps), \nabla\om(x_\eps) \bran,
\ee    
where $|\hat\ga|\leq|\ga|$.
Assume for definiteness that $\om(x_\eps) > 0$ and take $\ga:= -c_0 \eps |x_\eps|^{1-m_\om}$ with $c_0>0$ independent from $\eps$.
Then, using assumptions A3 and A2,  we find
\be\non
\begin{split}
\om(y_\ga) &\leq \eps|x_\eps|^{M_\om} - C_1c_0\eps|x_\eps|^{1-m_\om+2m_\om} + C_2 c_0^2\eps^2|x_\eps|^{2-2m_\om + 2(M_\om-1)} \blan x_\eps + \hat\ga\nabla\om(x_\eps)\bran^{M_\om-2} 
\\ &=
 \eps|x_\eps|^{M_\om}\big( 1 - C_1c_0 + C_2 c_0^2\eps|x_\eps|^{2-M_\om} \blan x_\eps + \hat\ga\nabla\om(x_\eps)\bran^{M_\om-2} \big). 
\end{split}
\ee
Since $(1-Cc_0\eps)|x_\eps|\leq | x_\eps + \hat\ga\nabla\om(x_\eps)| \leq (1+ Cc_0\eps)|x_\eps|$, 
for $c_0 = 2/C_1$ and $\eps\leq \eps_0(C, C_1,C_2,M_\om)$ sufficiently small we have $\om(y_\ga)<0$. 
Consequently, there is $|\ga'| < |\ga| =  c_0\eps |x_\eps|^{1-m_\om}$ such that $y_{\ga'}\in\Sigma$.
This gives the contradiction with the inclusion $x_\eps\in D_\kappa$ since  
$|x_\eps - y_{\gamma'}| < |\ga\nabla\om(x_\eps)| < \eps C |x_\eps| < \ka |x_\eps|$ if $\eps$ is sufficiently small but in the definition of $D_\kappa$ we have $m_\om + 2 - M_\om = 1$.

We have seen that $|\om(x)|\ge C(\ka)|x|^{M_\om}$ for $x\in D_\ka$. Then,
\be\non
\int_{D_\kappa} \frac{|F(x)|}{\om(x)^2}\, dx 
\le C_1(\ka) \int_{\R^d} \lan x\ran^{-M_F - 2M_\om} <\infty,
\ee
since $M_F>d-2M_\om$.
\qed

\subsection{Example}
\lbl{s:example}

In this section we apply Theorem~\ref{t:main} to the situation when the set $\Sigma$ is a quadric, a particular case of which was considered in \cite{K}.
Let 
\be\lbl{qqq}
q(x) = \frac12 x\cdot Bx  + a, \qquad x\in\R^d,
\ee
where $a\in \R$ and $B$ is a non-degenerate $d\times d$-matrix with $d\ge 2$ if $a\ne 0$ and $d\ge 4$ if $a=0$.
\footnote{The case when the quadratic polynomial $q$  contains a  linear term $l\cdot x$, $l\in\R^d$, reduces to the present one with $l=0$  via the transformation $x\mapsto x + B^{-1}l$.}
Consider the integral
\be\non
J_\nu = \int_{\R^d} \frac{G(x)}{q(x)^2 + \nu^2\Ga(x)^2}\, dx,
\ee
where the real-valued functions $G$ and $\Ga$ are $C^2$- and $C^4$-smooth, satisfying
\be\lbl{e:cond}
\begin{split}
|\p^\al G(x)| &\le C \lan x\ran^{-M_G - |\al|_1},  \qquad \forall\; 0\leq |\al|_1\leq 2, 
\\
\Ga(x) &\geq C^{-1}\lan x \ran^{r_*}, \qquad 
|\p^\al \Ga(x)| \le C \lan x\ran^{r_* - |\al|_1},  \qquad \forall\; 0\leq |\al|_1\leq 4. 
\end{split}
\ee
The real constants $M_G$ and $r_*$ satisfy
\be\lbl{e:cond_const}
M_G + r_* > d-2, \qquad M_G > d - 4.
\ee
Let 
$$
J_0(\nu) = 
\pi\int_{\Sigma_0} \frac{G(x)}{\Ga(x) |Bx| }\, d_\Sigma x,
$$
where $\Sigma$, $\Sigma_0$ and $d_\Sigma x$ are defined as before with the function $\om$ replaced by $q$.
\begin{corollary}
	\lbl{c:quard}
	 Under assumptions \eqref{e:cond} and \eqref{e:cond_const},
$	|J_\nu - \nu^{-1} J_0| \leq C\chi_{d,4}(\nu)$ if $a = 0$ and $d\geq 4$, and
$	|J_\nu - \nu^{-1} J_0| \leq C$ if $a \ne 0$ and $d\geq 2$.
\end{corollary}

In \cite{K} the corollary was proven in the case $d=2n$, $n\geq 2$, $a=0$ and
\be\lbl{ex:matr}B =\begin{pmatrix} 
	0 & \Id_{n\times n} \\
	\Id_{n\times n} & 0 
\end{pmatrix}.
\ee
Restrictions \eqref{e:cond_const} for the parameters $M_G, r_*, d$  coincide with those imposed in \cite{K}. 

\smallskip

{\it Proof of Corollary~\ref{c:quard}}. It suffices to check that assumptions A1-A5 above are satisfied for functions 
$F = G/\Ga^2$ and $\om = q/\Ga$.
A simple computation shows that \eqref{e:cond} implies assumptions A1-A3 with $M_F = M_G + 2r_*$, $M_\om = 2 - r_*$ and $m_\om = 1-r_*$.
Then, $M_{cr} = r_* - 2$ so \eqref{e:cond_const} implies assumption A4.
Assumption A5 follows from Lemma~\ref{l:A5}.
\qed

%
%
%
\subsection{Motivation}
\lbl{s:mot}

The wave turbulence theory was created in 1960-th to study small amplitude solutions of non-linear Hamiltonian PDE's with large space-period. Since then it has been intensively developing at heuristic level of rigour \cite{ZLF92, Naz11} while mathematical works, devoted to its rigorous justification, started to appear only several years ago (see \cite{LS, BGHS, DK, DKMV, DH1} and references therein). 
The central object in the wave turbulence is a non-linear kinetic equation, called the {\it wave kinetic equation}, 
that describes behaviour of certain physical characteristics of solutions to the underlying PDE. 
If the PDE describes the $N$-waves interaction, the $s$-th component, $s \in \R^n$, of the kinetic kernel $K$ is given by an integral of the form
\be\lbl{kin_int}
K_s = \int_{\R^{(N-1)n}}F_s(\xi,\s)\, \de^{\xi_1\dots\xi_p}_{\s_1\dots\s_{q-1}s} \,
\de\big(\om_{s}(\xi,\s)\big) \, d\xi_1\dots d\xi_p \, d\s_1\dots d\s_{q-1},
\ee 
where  $\xi_i,\s_j\in\R^n$ and $p+q = N$, see \cite[Sections 6.9, 6.11]{Naz11}. Here $\de^{\xi_1\dots\xi_p}_{\s_1\dots\s_{q-1}s}$ denotes the delta-function $\{\xi_1 +\dots+\xi_p = \s_1+\dots+\s_{q-1}+s\}$,
while $\de\big(\om_s\big)$ is the delta-function of 
\be\lbl{om:wt}
\om_s(\xi,\s)
= f(\xi_1) + \dots + f(\xi_p) - f(\s_1) - \dots f(\s_{q-1})- f(s),
\ee
where $f$ is the dispersion relation for the underlying PDE.
The delta-function $\de^{\xi_1\dots\xi_p}_{\s_1\dots\s_{q-1}s}$ means that the integration is performed over a hyper-space $\eqsim \R^{(N-2)n}$ while the subsequent multiplication by the delta-function  
$\de(\om_s)$ means that the integration is, in fact, taken over the set $\Sigma \subset \R^{(N-2)n}$ against the measure proportional to $\ds{\frac{d_\Sigma}{|\nabla \om_s|}}$, exactly as in \eqref{Sigma}, \eqref{asymp} with $\om = \om_s$, see \cite[p.~67]{ZLF92}.  
For rigorous mathematical treatment of delta-functions as $\de(\om_s)$ see \cite[Section~III.1.3]{Gelf}
or \cite[pp.~36-37]{Khin}.

A possible approach to study the wave turbulence rigorously is to add small viscosity and small random force to the PDE. In this setting nonlinearities of the form  \eqref{kin_int} and, in particular, the kinetic kernel $K$ above, appear as limits when $\nu\to 0$ of the integrals \eqref{set_1} with $\om$ given by \eqref{om:wt}, which makes   study of these limits crucial.   

Following this stochastic approach, in \cite{DK} we rigorously derived the wave kinetic equation for a quasisolution to the cubic NLS, describing the four-waves interaction.
In that case in \eqref{kin_int} we had $n\geq 2$, $p=q=2$ and $f(s) = |s|^2$. Then, expressing $\s_1$ via $\de^{\xi_1\xi_2}_{\s_1s}$ and denoting $x=\xi_1-s$, $y=\xi_2-s$, we got $\om_s(x,y) = -2 x\cdot y,$ where $x,y\in\R^n$, so $\om_s$ was a quadratic form given by the matrix proportional to \eqref{ex:matr} and $d=2n\ge 4$.
 By the explained above reason, analysis of the limiting as $\nu\to 0$ behaviour for the corresponding integral \eqref{set_1} played a crucial role for our research and was a subject of already cited separate publication \cite{K}.

As another example of equation to which our result applies let us consider the Petviashvilli equation, describing the three-waves interaction.
In the book \cite{Naz11} most of postulates of the wave turbulence are explained exactly on this example. In this case we have $n=2$, $p=2$, $q=1$ and $f(s) = s^1|s|^2$, where $s=(s^1,s^2)\in\R^2$.
Taking into account the relation $\xi_2 = s-\xi_1$ we get 
$\om_s(\xi_1) = s^1|\xi_1|^2 - 2(\xi_1^1-s^1) \xi_1\cdot s - \xi_1^1|s|^2$.
Then a straightforward computation shows that if $2s^1 \ne \pm |s|$, $\om_s$ can be written in the form \eqref{qqq} with $a\ne 0$, where $d=n=2$. 

For an example where the function $\om_s$ is not a quadratic polynomial, we suggest e.g. the Charney-Hassegawa-Mima equation giving a model for planetary Rossby waves and drift waves in inhomogeneous plasmas. This equation also describes the three-waves interaction but its dispersion relation has the form
$f(s) = -\beta\rho^2 s^1/(1+\rho^2|s|^2)$, where $\beta$ and $\rho$ are parameters of the system and again $s=(s^1,s^2)\in\R^2$, see \cite[Sections 7.6, 13.2.1]{Naz11}.
  
Mathematical theory of the wave turbulence has been intensively developing in the last several years and we believe that more wave systems will be rigorously studied  in the nearest future, using asymptotical expansion of integrals  like  \eqref{set_1}.   

\section{Manifold $\Sigma_0$ and its neighbourhood}
\lbl{s:neigh}

In this section we denote by $C, C_1,\dots$ various constants which  never depend on $\nu$ and $\Theta$,~\footnote{The parameter $\Theta$ will be introduced below and should be sufficiently small.} and may change from line to line.

Recall that we give the proof in the case $d\geq 4$, assuming that $x=0$ is the unique critical point of the function $\om$ on the set $\Sigma\cap \supp F$.

\subsection{Neighbourhood of the manifold $\Ss$}
Let us choose differentiable coordinates $\xi\in\R^{d-1}$ on the manifold $\Ss$ and consider the coordinate functions $\xi\mapsto x_\xi\in\Ss.$ 
Abusing notation we will often write $\xi\in\Ss$. Denote
$$
N_\xi = \nabla\om (x_\xi).
$$ 
The vector $N_\xi$ is orthogonal to $\Sigma_0$ at the point $\xi$.
Consider the mapping $\pi:\Ss\times\R \mapsto \R^d$ defined as
\be\lbl{pi}
\pi(\xi, \theta) = x_\xi + \theta N_\xi.
\ee
Let $0<\Theta\leq 1$ and
\be\non
U_\Theta(\Sigma_0) = \{x = \pi(\xi,\theta):\, \xi\in\Ss, \; |\theta| <  \theta_\xi(\Theta)\},
\ee
where 
\be\lbl{theta_xi}
\theta_\xi = \theta_\xi(\Theta) = \frac{\Theta}{\lan x_\xi\ran^{M_\om - 2}}
\ee
and we recall that $M_\om$ is defined in assumption A2.
Below we assume $\Theta$ to be sufficiently small but fixed and independent from $\nu$. 

Let us denote the Hessian
$$
\Om(x) = \operatorname{Hess} \om(x)
$$
and by $\|\Om\|$ denote the operator norm of $\Om$.
Assumption A2 implies the following result.
\begin{lemma}\lbl{l:prem}
(1) There is a constant $C>0$ such that for any $\xi\in\Sigma_0$, 
\be\lbl{N_xi_upp_est}
(a)\qu \theta_\xi |N_\xi| \le C \Theta |x_\xi | \qu\qnd\qu
(b)\qu \theta_{\xi}  \|\Om(x_\xi)\| \le C\Theta. 
\ee

(2) 	There is a constant $C>0$ such that for any $x\in U_\Theta(\Sigma_0)$, given by  $x=\pi(\xi,\theta) = x_\xi+ \theta N_\xi$,
\be\non
(1-C\Theta)|x_\xi| \leq |x|\leq (1+C\Theta)|x_\xi|.
\ee
\end{lemma}
{\it Proof. (1)} 
Assumption A2 and \eqref{theta_xi} immediately imply
the estimate {\it (1b)} and the relation
$
\theta_\xi |N_\xi| \le C \Theta \lan x_\xi \ran.
$ The latter implies {\it (1a)} for $|x_\xi|\ge 1$,
while for $|x_\xi|< 1$ it follows from the relation 
$\nabla\om(0) = 0$ and $\theta_\xi = \Theta$.
 
{\it (2)} In view of \eqref{N_xi_upp_est}(a),
$
|x|\leq |x_\xi| + \theta_\xi|N_\xi| \leq (1+C\Theta) |x_\xi|
$
and
$
|x_\xi| \leq |x| + \theta_\xi|N_\xi| \leq |x| + C\Theta|x_\xi|,
$
which imply the desired inequality.
\qed

\smallskip

Now we show that $(\xi,\theta)$ from coordinates in the set $U_\Theta(\Sigma_0)$ and write the function $\om$ in these coordinates.

\begin{proposition}
	(1) For any sufficiently small $\Theta$ the set $U_\Theta(\Sigma_0)$ is uniquely parametrized by coordinates $\big\{(\xi,\theta):\; \xi\in\Ss,  \, |\theta| < \theta_\xi \big\}$ via formula \eqref{pi}.
	
	(2) Let $x = \pi(\xi,\theta)\in U_\Theta(\Sigma_0)$. Then
	\be\lbl{om}
	\om(x) = \theta |N_\xi|^2 g_\xi(\theta),
	\ee
	where the function $\theta\mapsto g_\xi(\theta)$ is $C^2$-smooth and satisfies $g_\xi(0) = 1$, 
	\be\lbl{g_xi}
	|g_\xi(\theta)-1|\leq C\Theta, \qquad 
	|g_\xi'(\theta)|\leq C\lan x_\xi \ran^{M_\om - 2},\qquad
	|g_\xi''(\theta)|\leq C\lan x_\xi\ran^{2(M_\om - 2)}.
	\ee
\end{proposition} 
	{\it Proof.}
	{\it (1)} We need to show that if $\Theta$ is sufficiently small, then for any $\xi_{i}\in\Ss$ and $|\theta_{i}|<\theta_{\xi_i}(\Theta)$, $i=1,2$, satisfying $\pi(\xi_1,\theta_1)=\pi(\xi_2,\theta_2)$ we have $x_{\xi_1}=x_{\xi_2}$ and $\theta_1=\theta_2$.
	From \eqref{pi} we get 
	\be\lbl{x-x}
	x_{\xi_1} - x_{\xi_2} = \theta_2 N_{\xi_2} - \theta_1N_{\xi_1}.
	\ee
	Taking the scalar product of the both sides of this equation with $x_{\xi_1} - x_{\xi_2}$ we get
	\be\lbl{del_x}
	|x_{\xi_1} - x_{\xi_2}|^2 = \theta_2\lan N_{\xi_2}, x_{\xi_1} - x_{\xi_2}\ran -
	\theta_1\lan N_{\xi_1}, x_{\xi_1} - x_{\xi_2}\ran.
	\ee
	Recall that $N_{\xi_i} = \nabla \om(x_{\xi_i})$. Since $\om(x_{\xi_1}) = \om(x_{\xi_2}) = 0$, the Taylor formula applied to the function $\om$ at the point $x_{\xi_2}$ implies
	\be\non
	\big|\lan N_{\xi_2}, x_{\xi_1} - x_{\xi_2}\ran \big| \leq \frac12 \max\limits_{x\in [x_{\xi_1}, x_{\xi_2}]} 
	\|\Om(x)\||x_{\xi_1} - x_{\xi_2}|^2,
	\ee
	where $[x_{\xi_1}, x_{\xi_2}]$ denotes the segment in $\R^d$ connecting $x_{\xi_1}$ and $x_{\xi_2}$.
	Clearly, the same estimate holds for the scalar product $\lan N_{\xi_1}, x_{\xi_1} - x_{\xi_2}\ran$.
	Thus, \eqref{del_x} implies 
	\be\lbl{xi1-xi2}
	|x_{\xi_1} - x_{\xi_2}|^2  \le \frac{\theta_{\xi_1} + \theta_{\xi_2}}{2} \max\limits_{x\in [x_{\xi_1}, x_{\xi_2}]} \|\Om(x)\||x_{\xi_1} - x_{\xi_2}|^2.
	\ee
	We claim that for any  $x\in [x_{\xi_1}, x_{\xi_2}]$ and $i=1,2$
	\be\lbl{xi12}
	C^{-1}|x_{\xi_i}| \le | x| \le C | x_{\xi_i}|.
	\ee
	Together with assumption A2 this would imply that the r.h.s. of \eqref{xi1-xi2} is bounded by
	$ C\Theta|x_{\xi_1} - x_{\xi_2}|^2$, so for  $\Theta < C^{-1}$ we get $x_{\xi_1} = x_{\xi_2}$. Consequently, $\theta_1=\theta_2$, so the proof is finished.
	
	It remains to establish \eqref{xi12}. In view of \eqref{N_xi_upp_est}(a), \eqref{x-x} implies
	\be\lbl{xi_diff}
	|x_{\xi_1} - x_{\xi_2}| \le |\theta_2N_{\xi_2}| + |\theta_1 N_{\xi_1}| \le C\Theta( |x_{\xi_1}| + |x_{\xi_2}|),
	\ee
	so 
	\be\lbl{equiv_24}
	C^{-1}|x_{\xi_2}| \le | x_{\xi_1}| \le C | x_{\xi_2}|, 
	\ee
	if $\Theta$ is sufficiently small.
	Assume for definiteness that $|x_{\xi_2}| \geq |x_{\xi_1}|$. 
	Any $x\in [x_{\xi_1}, x_{\xi_2}]$ satisfies
	$$|x_{\xi_2}| - |x_{\xi_1} - x_{\xi_2}| \leq |x| \leq |x_{\xi_1}| + |x_{\xi_2}|.$$
	Then, by \eqref{xi_diff} and \eqref{equiv_24} we get
	$C^{-1}|x_{\xi_2}| \leq |x| \leq C|x_{\xi_1}| $, so \eqref{xi12} holds true.

	\smallskip
	
	{\it (2)} We apply the Taylor formula at $\theta=0$ to the function $\theta\mapsto \om_\xi(\theta):= \om\big(\pi(\xi,\theta)\big)$. 
	To this end we compute the derivatives
	$$
	\om_\xi'(0) = \frac{d}{d\theta}\Big|_{\theta=0} \omega(x_\xi + \theta N_\xi) 
	=\lan \nabla\om(x_\xi), N_\xi\ran = |N_\xi|^2
	$$
	and
	$
	\om_\xi''(\theta) = \lan\Om(x_\xi + \theta N_\xi)N_\xi,N_\xi \ran.
	$
	Since $\om_\xi(0) = 0$, we find
	\be\non
	\om_\xi(\theta) = \theta |N_\xi|^2 + \int_0^\theta \lan \Om(x_\xi + tN_\xi)N_\xi, N_\xi\ran (\theta - t)\, dt.
	\ee
	Thus, we get \eqref{om} with 
	\be\non
	g_\xi(\theta) = 1 + \int_0^\theta \lan \Om(x_\xi + tN_\xi)n_\xi, n_\xi\ran \frac{\theta - t}{\theta}\, dt
	=1 + \theta\int_0^1 \lan \Om(x_\xi + s\theta N_\xi)n_\xi, n_\xi\ran (1 - s)\, ds,
	\ee	
	where $n_\xi = N_\xi / |N_\xi|$.
	Since $\om$ is $C^4$-smooth,  the function $g_\xi$ is $C^2$-smooth. 
	
	Let $B_\xi = [x_\xi - \theta_\xi N_\xi,\, x_\xi + \theta_\xi N_\xi]$.
	Using assumption A2 and then Lemma~\ref{l:prem}(2), we get
	\be\non
	|g_\xi(\theta) - 1| \leq \theta_\xi \max_{x\in B_\xi} \|\Om(x)\| 
	\le C \frac{\Theta}{\lan x_\xi\ran^{M_\om-2}}\max_{x\in B_\xi}\lan x \ran^{M_\om - 2} \le C_1\Theta,
	\ee
	which is the first inequality from \eqref{g_xi}.
	Next, 
	\be\non
	g'_{\xi}(\theta)=\int_0^1 \big\lan \Om(x_\xi + s\theta N_\xi)n_\xi, n_\xi\big\ran (1 - s)\, ds
	+ \theta\int_0^1 \big\lan \sum_{i = 1}^d \p_{x_i}\Om(x_\xi + s\theta N_\xi)n_\xi, n_\xi\big\ran 
	 (N_\xi)_i\, s(1 - s)\, ds.
	\ee
	Using again assumption A2 and Lemma~\ref{l:prem}(2), we find
	\be\non
	|g'_{\xi}(\theta)| \leq C\Big(\lan x_\xi\ran^{M_\om-2} 
	+ \frac{\Theta}{\lan x_\xi\ran^{M_\om-2}} 
	\lan x_\xi \ran^{M_\om-3}\lan x_\xi \ran^{M_\om-1} \Big) = C_1\lan x_\xi \ran^{M_\om-2},
	\ee
	which is the second inequality in \eqref{g_xi}. The third one
follows similarly.
	\qed
	
	\smallskip
	
	Next we write the volume element $dx$ in the coordinates $(\xi,\theta)$.
	\begin{proposition}\lbl{l:volume}
	In the set $U_\Theta(\Sigma_0)\subset \R^d$ the volume element $dx$, written in the coordinates $(\xi,\theta)$, takes the form 
	\be\lbl{dx=}
	dx = |N_\xi|\,\mu_\xi(\theta)\,d\theta\, m(d\xi),
	\ee
	where $m(d\xi)$ denotes the volume element on $\Sigma_0$ and the density function $\theta\mapsto \mu_\xi(\theta)$ is  a polynomial of degree $d-1$ satisfying 
	\be\lbl{mu_prop}
	\mu_\xi(0) = 1,  \qu \mu_\xi(\theta)\ge C^{-1}>0,
	\quad \big|\frac{d^p}{d\theta^p} \mu_\xi(\theta)\big|\le C\lan x_\xi\ran^{p(M_\om-2)},
	\ee
	for all $\xi\in\Sigma_0$, $|\theta|< \theta_\xi$ and $0\leq p\leq d-1$, if $\Theta$ is sufficiently small.
	\end{proposition}
	{\it Proof.} Consider the projection 
	$\Pi: U_\Theta(\Sigma_0) \mapsto\Sigma_0$ mapping $x=\pi(\xi,\theta)$ to $x_\xi= \pi(\xi, 0)$,
	\be\lbl{prof_P}
	\Pi(x) = x_\xi .
	\ee
	Kernel of the differential $d\Pi(x)$, taken at a point $x=\pi(\xi, \theta)$,  is given by the linear span of the vector $N_\xi$. 
	Let $J_\Pi(x)$ be the Jacobian of the linear map $d\Pi(x)$, restricted to the orthogonal complement $\big(\operatorname{span}(N_\xi)\big)^\perp$. Accordingly to the coarea formula \cite[Section 13.4]{BZ}, for any Lebesgue measurable set $A\subset U_\Theta(\Sigma_0)$,
	\be\non
	\int_A J_\Pi(x)\,dx = \int_{\Sigma_0} Leb\big(A\cap \Pi^{-1}(x_\xi)\big)\, dm(\xi),
	\ee 
	where $dm(\xi)$ denotes the volume element on $\Sigma_0$ and $Leb(\cdot)$ stands for the Lebesque measure on the set 
	$\Pi^{-1}(x_\xi) = \big\{x=\pi(\xi,\theta): \, |\theta|\leq \theta_\xi\big\}$.
	Since $Leb\big(A\cap \Pi^{-1}(x_\xi)\big) = |N_\xi| Leb\big(\{\theta:\,\pi(\xi,\theta)\in A\}\big)$, we find
	$
	J_\Pi(x) \,dx = |N_\xi| \, d\theta\,dm(\xi).
	$
	Thus, \eqref{dx=} takes place with 
	\be\non
	\mu_\xi(\theta) = \frac{1}{J_\Pi(\xi,\theta)}, \qmb{where} \qu J_\Pi(\xi,\theta):=J_\Pi\big(\pi(\xi,\theta)\big).
	\ee
	Analysis of  the density function $\mu$ relies on the following lemma.
	\begin{lemma}\lbl{l:dPi}
		We have
		\be\non
		J_\Pi (\xi,\theta) = \Big(\det\big(\Id_{d-1} + \theta\Om_{d-1}(x_\xi)\big)\Big)^{-1},
		\ee
		where $\Id_{d-1}$ denotes the $(d-1)\times(d-1)$-identity matrix 
		and $\Om_{d-1}(x_\xi)$ is the matrix obtained from the Hessian $\Om(x_\xi)$ by deleting its last row and column, where $\Om(x_\xi)$ is written with respect to any orthonormal basis in which the last vector coincides with $N_\xi/\|N_\xi\|$.
	\end{lemma}
{\it Proof.} We fix $x=\pi(\xi,\theta)$, a vector $v\perp N_\xi$ and take $t\in\R$ so small that $x+vt \in U_\Theta(\Sigma_0)$.  
Then $x+vt = \pi(\xi(t), \theta(t))$ for appropriate $\xi(t)$ and $\theta(t)$,  satisfying $\xi(0)=\xi$ and $\theta(0) = \theta$. 
In other words, $\pi(\xi,\theta) +vt = \pi(\xi(t), \theta(t))$, or,  in more details,
\be\non
\Pi(x) + \theta N_\xi + tv = \Pi(x+tv) + \theta(t)N_{\xi(t)}.  
\ee
Differentiating the last equation with respect to $t$ at the point $t=0$, we get
\be\lbl{dPi}
 v = d\Pi(x)v + \theta'(0) N_\xi  + \theta (N_{\xi(t)})'_{t=0}. 
\ee
Next,  
$(N_{\xi(t)})'_{t=0}  = \ds{\frac{d}{dt}\Big|_{t=0} \nabla \om(\Pi(x+tv)) = \Om(x_\xi)\big(d\Pi(x)v\big)}$. 
Since the vectors $v$ and $d\Pi(x)v$ are orthogonal to $N_\xi$, taking the projection $\Pr$ on the space $N_\xi^{\perp}\eqsim \R^{d-1}$ in \eqref{dPi} we get
\be\non
\big(\Id_{d-1} + \theta \Pr\circ\Om(x_\xi)\big)d\Pi(x)v = v, 
\ee
where  the vectors $v, d\Pi(x)v \in N_\xi^{\perp}$ are viewed as $d-1$-vectors.
In the basis as in the formulation of the lemma we have $ \Pr\circ\Om(x_\xi) = \Om_{d-1}(x_\xi)$.
In view of \eqref{N_xi_upp_est}(b), the operator $\Id_{d-1} + \theta \Om_{d-1}(x_\xi)$ is invertible if $\Theta$ 
 is sufficiently small, so we get
 $$
 d\Pi(x) = \big(\Id_{d-1} +\theta \Om_{d-1}(x_\xi)\big)^{-1} \qnd J_\Pi(\xi,\theta) = \frac{1}{\det\big(\Id_{d-1} +\theta \Om_{d-1}(x_\xi)\big)}.
 $$
 \qed 
	
By Proposition~\ref{l:dPi}, 
\be\non
	\mu_\xi(\theta) = \det\big(\Id_{d-1} + \theta\Om_{d-1}(x_\xi)\big).
\ee
Clearly, $\mu_\xi(0) = 1$. From the relation \eqref{N_xi_upp_est}(b) we get the lower bound $\mu_\xi(\theta) \geq C > 0$ uniformly in $\xi\in\Sigma_0$ and $|\theta|< \theta_\xi$, if $\Theta$ is sufficiently small.
Thus, it remains to get the estimate for derivatives of $\mu_\xi$.
Let us write 
\be\non
\mu_\xi(\theta) = \sum_{k=0}^{d-1} \theta^k P_k(x_\xi) 
\ee
where the function $P_l(x_\xi)$ is a homogeneous polynomial of degree $l$ with respect to the second derivatives $\p_i\p_j \om(x_\xi)$ (in particular, $P_0(x_\xi)=1$ and $P_1(x_\xi) = \operatorname{tr}\Om_{d-1}(x_\xi)$).
Then, $\ds{\frac{d^p}{d\theta^p}\mu_\xi(\theta) = \sum_{k=0}^{d-p-1}C_{k,p}\theta^k P_{k+p}(x_\xi)}$,
so, by assumption A2, 
$$
\big|\frac{d^p}{d\theta^p}\mu_\xi(\theta)\big| \leq C \max_{0\leq k \leq d-p-1} (\theta_\xi)^k \lan x_\xi\ran^{(k+p)(M_\om-2)} \leq C_1\lan x_\xi\ran^{p(M_\om-2)}, 
$$
according to \eqref{theta_xi}.
\qed

\smallskip

We conclude this section with the following characterization of the set~$U_\Theta(\Sigma_0)$.
\begin{lemma}\lbl{l:comp_neig}
	Let $\ka = \ka(\Theta)$ be sufficiently small. Then any $x\in\R^d$ satisfying 
	\be\lbl{comp_neig}
	| x - \Sigma_0| \leq \ka\min\big( |x|, |x|^{m_\om  + 2 - M_\om} \big)
	\ee
	belongs to $U_\Theta(\Sigma_0)$.
\end{lemma}

{\it Proof of Lemma~\ref{l:comp_neig}.}
We start by claiming that for any $x$ satisfying \eqref{comp_neig} 
\be\lbl{dist_ach}
|x-\Sigma_0| = |x-x_\xi| \qmb{for some}\qu \xi\in\Sigma_0.
\ee
Indeed, since the set $\Sigma$ is closed, $|x-\Sigma| = |x-y|$ for some $y\in\Sigma$.
If $y\in\Sigma_0$ then the claim is true. Otherwise $y=0$, so $|x-\Sigma_0| = |x|$. But then $x$ does not satisfy \eqref{comp_neig} once $\ka < 1$. 

Relation \eqref{dist_ach} implies that for any $x$ satisfying \eqref{comp_neig}, there is $\xi\in\Sigma_0$ such that
\be\lbl{x-x-xi}
x = x_\xi+ \theta N_\xi \qmb{where $\theta$ satisfies} \qu |\theta N_\xi| \leq \ka\min(|x|, |x|^{m_\om  + 2 - M_\om}). 
\ee
It remains to show that $|\theta|< \theta_\xi$.
To this end we note that, according to \eqref{x-x-xi},
\be\lbl{x-x-xi1}
C^{-1}|x_\xi| \leq |x| \leq C|x_\xi|, \qmb{uniformly in}\qu 0< \ka \leq 1/2.  
\ee
Assume first that $|x| \geq 1$. 
Then $\min\big(|x|, |x|^{m_\om  + 2 - M_\om}\big) = |x|^{m_\om  + 2 - M_\om}$, by \eqref{M-m}. 
So, by \eqref{x-x-xi} and \eqref{x-x-xi1},
$|\theta N_\xi| \leq C\ka |x_\xi|^{m_\om  + 2 - M_\om}.$ 
Then, using assumption A3,  we find
$|\theta|\leq  C\ka|x_\xi|^{2 - M_\om} < \theta_\xi$ if $\ka < C^{-1}\Theta.$
In the case $|x|\leq 1$,  
we use the inequality $|\theta N_\xi| \leq C\ka|x_\xi|$, following from \eqref{x-x-xi}, \eqref{x-x-xi1}. 
Due to \eqref{N_xi-lower}, it implies
$|\theta|\leq C\kappa < \theta_\xi$ if $\kappa< C^{-1}\Theta$. 
\qed

\begin{corollary}\lbl{c:int_over_comp}
Assumption A5 implies that
$
\ds{\int_{B_1^c\sm U_\Theta(\Sigma_0)} \frac{|F(x)|}{\om(x)^2} \,dx  < \infty.}
$
\end{corollary}
{\it Proof.}
According to Lemma~\ref{l:comp_neig}  together with \eqref{M-m}, the domain $B^c_1\sm U_\Theta(\Sigma_0)$ is contained in the set $D_\kappa$ from assumption A5, if $\kappa$ is sufficiently small.
\qed

\subsection{Integrals over the manifold $\Sigma_0$ }

Recall that $B_r$ denotes a closed ball in $\R^d$ of radius $r$ centred at zero and  set 
\be\lbl{R_ab}
R_a^b = B_b\sm B_a, \qquad 0<a<b.
\ee
Recall also the notation \eqref{chi}.
\begin{lemma}\lbl{l:int_zero} 
(1)	Assume that $\R\ni n \leq d -1$. Then for any $0 <\de \le 1/2$,
	\be\lbl{8521}
	\int_{\Sigma_0\cap R_\de^1} |x_\xi|^{-n} \, m(d\xi)  \le C\, \chi_{d, n+1}(\de).
	\ee
	If $n<d-1$, then 
	$\int_{\Sigma_0\cap B_\de} |x_\xi|^{-n} \, m(d\xi) \leq C \de^{d-1-n}$. In particular, $\int_{\Sigma_0\cap B_1} |x_\xi|^{-n}\, m(d\xi)<\infty$. 
	
(2) Assume that $\R\ni n> M_\om - m_\om -2 + d$.
	Then
	$$\int_{\Sigma_0\sm B_1} |x_\xi|^{-n} \, m(d\xi)<\infty.$$ 
\end{lemma}
{\it Proof.}
Let us write 
$\ds{m(d\xi) = \frac12\int_{-\theta_\xi}^{\theta_\xi} \frac{|N_\xi|\mu_\xi(\theta)\,d\theta m(d\xi)}{\theta_\xi|N_\xi|\mu_\xi(\theta)}}$. Then,
using \eqref{dx=}, for any $A\subset\Sigma_0$
we find
\be\non	
J:= \int_{A} |x_\xi|^{-n} \, m(d\xi) 
= 	\frac12\int_{\Pi^{-1}(A)} \frac{|x_\xi|^{-n}}{\theta_\xi |N_\xi|\mu_\xi(\theta)} \, dx,
\ee
where $(\xi,\theta) = \pi^{-1}(x)$ and the projection $\Pi$ is defined in \eqref{prof_P}.

{\it (1)} To establish item (1) we set $A = \Sigma_0\cap R_\de^1$.
By Lemma~\ref{l:prem}(2), $\Pi^{-1}(\Sigma_0\cap R^1_\de)\subset R_{c_0\delta}^{c_1}$ for appropriate constants $c_0, c_1>0$. Then, using that $\theta_\xi \geq C\Theta$ for $x_\xi \in R_{c_0\delta}^{c_1}$, \eqref{N_xi-lower} and that $\mu_\xi(\theta)\geq C$, we get
\be\non
J \leq C\Theta^{-1}\int_{ R_{c_0\delta}^{c_1}} |x_\xi|^{-n-1} \, dx 
\leq C_1 \Theta^{-1}\int_{ R_{c_0\delta}^{c_1}} |x|^{-n-1} \, dx 
= C_1\Theta^{-1}\int_{c_0\de}^{c_1} r^{-n+d-2}\,dr,
\ee
which implies the first assertion of item (1).
\footnote{The constant $C$ in \eqref{8521} does not depend on $\Theta$ since the integral in \eqref{8521} is independent from $\Theta$ (and the argument above works with any sufficiently small but fixed $\Theta$).}
The second assertion follows by taking $A = \Sigma_0\cap B_\de$ and replacing $R_{c_0\de}^{c_1}$ by $B_{c\de}$ in the formula above. 

\smallskip

{\it (2)} 
Now we set $A = \Sigma_0\sm B_1$. Since $\Pi^{-1}(\Sigma_0\sm B_1)\subset B_{r}^c$ with some $r> 0$, by assumption A3 and \eqref{theta_xi} we find
\be\non
J \leq C\Theta^{-1}\int_{ B_{r}^c} |x|^{-n + M_\om - 2 -m_\om} \, dx < \infty
\ee
once $-n + M_\om - 2 -m_\om < -d $.
\qed

\begin{corollary}\lbl{c:integration} 
Let $g:\Sigma_0\mapsto\R$ be a measurable function. 
Then the integral
$\ds{\int_{\Sigma_0} \frac{g(x_\xi)}{|N_\xi|^k}\, m(d\xi)}$
with $0\leq k<3$ converges if
$|g(y)|\leq C\lan y\ran^{-n}$ with
\be\lbl{nnn}
n> M_\om - (k+1) m_\om -2 + d.
\ee  
If  \eqref{nnn} holds with $k=3$, then
$$
\Big|\int_{\Sigma_0\sm B_\de} \frac{g(x_\xi)}{|N_\xi|^3}\, m(d\xi)\Big|
\leq
C\chi_{d,4}(\de) \qmb{for}\qu 0<\de\leq 1/2.
$$	
\end{corollary}
{\it Proof.} The assertion immediately follows from Lemma~\ref{l:int_zero}, assumption~A3 and \eqref{N_xi-lower} together with the restriction $d\ge 4$. 
\qed

\section{Integral $I_\nu$}
\lbl{s:the_integral}

In this section we prove Theorem~\ref{t:main}.
We fix $\Theta$ sufficiently small and allow the constants $C, C_1,\dots$ to depend on $\Theta$ (but not on $\nu$).
For a domain $W\subset\R^d$ we denote by $\lan I_\nu, W \ran$ the integral $I_\nu$ taken not over $\R^d$ but over $W$: 
$$
\blan I_\nu, W \bran := \int_W \frac{F(x)}{\om(x)^2 + \nu^2}\, dx.
$$
We set 
\be\lbl{de-al}
\de := \al\sqrt \nu,
\ee
where $\al=\al(\Theta)> 0$ is a $\nu$-independent constant, which we will choose later (typically it is a big constant).
We write the integral $I_\nu$ as the sum
\be\lbl{I_nu_sum}
I_\nu = \blan I_\nu, B_\de \bran + \blan I_\nu, B_\de^c\sm U_\Theta(\Sigma_0) \bran + \blan I_\nu, U_\Theta(\Sigma_0) \sm B_\de \bran
\ee
and show that the sum of the first two integrals is bounded by $C\chi_{d,4}(\nu)$ while the third one leads to the desired asymptotic.

\subsection{The first and second integrals in \eqref{I_nu_sum}}

For the first integral in \eqref{I_nu_sum} we use the trivial estimate 
\be\lbl{first_int}
\big|\blan I_\nu, B_\de \bran\big| \leq C\de^{d}\nu^{-2}  = C\al^d\nu^{d/2 - 2} \leq C_1,
\ee
since $d\geq 4.$
To analyse the second integral, we take $0<\hat\de<1$ sufficiently small but independent from $\nu$ (see below),  
 and write
\be\non
\blan I_\nu, B_\de^c\sm U_\Theta(\Sigma_0) \bran  
= \blan I_\nu, B_1^c\sm U_\Theta(\Sigma_0)\bran +  \blan I_\nu, R_{\hat\de}^1\sm U_\Theta(\Sigma_0)  \bran
+ \blan  I_\nu, R_\de^{\hat\de}\sm U_\Theta(\Sigma_0) \bran,
\ee
where we recall notation \eqref{R_ab}.
According to Corollary~\ref{c:int_over_comp}, the first summand above is bounded by a constant.
Since the set $R_{\hat\de}^1\sm U_\Theta(\Sigma_0)$ is separated from $\Sigma_0$, is bounded and independent from $\nu$, the same bound is true for the second summand as well. 

To analyse the third summand we employ Lemma~\ref{l:comp_neig}. By \eqref{M-m}, for $x\in R_\de^{\hat\de}$
the r.h.s. of \eqref{comp_neig} equals $\ka|x|$. Then,
\be\lbl{neig-mid}
\big|\blan  I_\nu, R_\de^{\hat\de}\sm U_\Theta(\Sigma_0) \bran\big|
\leq
C\int\limits_{\{x\in R_\de^{\hat\de}:\, |x-\Sigma_0| > \ka|x|\}} \frac{dx}{\om(x)^2}.
\ee
According to the Morse lemma, for sufficiently small $r>0$ there is a $C^2$-diffeomorphism $x\mapsto y$, $B_{r} \mapsto y(B_r)$, 
such that $y(0) = 0$ and the function $q(y) = \om(x(y))$ is a non-degenerate quadratic form.
Let us choose $\hat\de, \hat r > 0$ in such a way that $$y(B_{\hat \de})\subset B_{\hat r}\subset y(B_{r}).$$
Clearly, the set $\{x\in R_\de^{\hat\de}:\, |x-\Sigma_0| > \ka|x|\}$ is contained in the set 
$A:=\{x\in R_\de^{\hat\de}:\, |x-\Sigma\cap B_r| > \ka|x|\}$.
Since $C^{-1}|x_1 - x_2| \leq |y(x_1) - y(x_2)|\leq C|x_1 - x_2|$ for any $x_1, x_2 \in B_r$, image of the set $A$  under the mapping $x\mapsto y$ is contained in the set 
$\{y\in R_{c\de}^{\hat r}:\, |y-y(\Sigma\cap B_r)| > c\ka|y|\}$ for appropriate $c<1$. In turn, this set is contained in 
the set 
$A^y:=\{y\in R_{c\de}^{\hat r}:\, |y-y(\Sigma)\cap B_{\hat r}| > c\ka|y|\}$. 
Since the set $y(\Sigma)\cap B_{\hat r}$ is a cone $\Sigma^q = \{y\in\R^d:\, q(y) = 0\}$ intersected with the ball $B_{\hat r}$, 
$
A^y = \{y\in R_{c\de}^{\hat r}:\, |y-\Sigma^q| > c\ka|y|\}.
$
Since on the set  $ A^y \cap \{|y| = \hat r\}$ the quadratic form $q$ is separated from zero and $\Sigma^q$ is a cone, we have $|q(y)|\geq C |y|^2$ for any $y \in A^y $. 
Then, the r.h.s. of \eqref{neig-mid} is bounded by
$$
\ds{C\int_{A^y}\frac{dy}{q(y)^2} \leq C_1\int_{c\de}^{\hat r} \frac{r^{d-1}\,dr}{r^4} \leq C_2\chi_{d,4}(\de).}
$$
Thus, we have seen that
\be\lbl{second_int}
\big|\blan  I_\nu, B_\de^c\sm U_\Theta(\Sigma_0) \bran\big|
\leq
 C\chi_{d,4}(\de).
\ee

\subsection{The third integral in \eqref{I_nu_sum}: first approximations}

{\it Step 1.} Consider the set 
$$
V_\de =\big\{x = \pi(\xi,\theta)\in U_\Theta(\Sigma_0):\, x_\xi\in\Sigma_0\sm B_\de, \, |\theta|\leq \theta_{\xi}\big\}.
$$
Using Lemma~\ref{l:prem}(2) it is straightforward to see that
the set $V_\de \, \Del \,
(U_\Theta(\Sigma_0)\sm B_\de)$ is contained in a ball $B_{c\de}$ for appropriate constant $c\geq 1$.
Then, arguing as in \eqref{first_int}, we see that
\be\lbl{third_int_appr}
\big|\blan I_\nu, V_\de \bran - \blan I_\nu, U_\Theta(\Sigma_0)\sm B_\de \bran  \big| \leq C_1.
\ee
Thus, it remains to study the integral $\blan I_\nu, V_\de \bran$. 
According to Proposition~\ref{l:volume}, written in the $(\xi,\theta)$-coordinates it takes the form
$$ 
\blan I_\nu, V_\de \bran = \int_{\Sigma_0\sm B_\de} m(d\xi)\,\int_{-\theta_\xi}^{\theta_\xi} d\theta\,\frac{|N_\xi|\mu_\xi(\theta) F(\xi,\theta)}{\om^2(\xi,\theta) + \nu^2}.
$$
Using \eqref{om}, we rewrite the integral as
\be\lbl{fin_form}
\blan I_\nu, V_\de \bran  = \int_{\Sigma_0\sm B_\de} \frac{m(d\xi)}{|N_{
	\xi}|^3}\,\int_{-\theta_\xi}^{\theta_\xi} \frac{\Phi(\xi,\theta)}{\theta^2g^2_\xi(\theta) + \eps_\xi^2}\,d\theta,
\ee
where
$$
\Phi(\xi,\theta):= \mu_\xi(\theta) F(\xi,\theta)
\qnd
\eps_\xi:=\nu |N_\xi|^{-2}.
$$

{\it Step 2. } 
Consider the internal integral in \eqref{fin_form}
$$
J_\nu(\xi): = \int_{-\theta_\xi}^{\theta_\xi} \frac{\Phi(\xi,\theta)}{\theta^2g^2_\xi(\theta) + \eps_\xi^2}\,d\theta
$$
and denote by
$J_\nu^0(\xi)$ the integral  $J_\nu(\xi)$ in which the functions $\Phi(\xi,\cdot)$ and $g_\xi$ are replaced by their values at zero  $\Phi(\xi,0) = F(\xi,0)$ and $g_\xi(0)=1$:
\be\lbl{J_nu^0}
J_\nu^0(\xi): = \int_{-\theta_\xi}^{\theta_\xi} \frac{F(\xi,0)}{\theta^2+ \eps_\xi^2}\,d\theta
= 2\frac{F(\xi,0)}{\eps_\xi}\arctan \frac{\theta_\xi}{\eps_\xi}.
\ee
In this step we will show that it suffices to analyse the integral \eqref{fin_form} in which the internal integral $J_\nu(\xi)$ is replaced by $J_\nu^0(\xi)$. To this end for fixed $\xi$ and $\theta$ we consider the function
\be\non
f_{\xi,\theta}(t) := \frac{\Phi(\xi,t)}{\theta^2g^2_\xi(t)+ \eps_\xi^2}.
\ee
By the Taylor formula,
$$
f_{\xi,\theta}(t) - f_{\xi,\theta}(0) = f_{\xi,\theta}'(0)t
+ \frac12 f_{\xi,\theta}''(\hat t(t;\xi,\theta))t^2,
$$
where $|\hat t|\leq |t|$.
Since $f'_{\xi,\theta}(0) = f'_{\xi,-\theta}(0)$,
\be\lbl{2309}
J_\nu(\xi) - J_\nu^0(\xi) = \int_{-\theta_\xi}^{\theta_\xi}
\big(f_{\xi,\theta}(\theta) - f_{\xi,\theta}(0)\big) \, d\theta
=   \int_{-\theta_\xi}^{\theta_\xi}\frac{\theta^2}{2} f_{\xi,\theta}''(\hat t) \,d\theta.
\ee
Next we estimate the derivative $f_{\xi,\theta}''$.
Since for $x=\pi(\xi,t)$ we have $\p_{t} = N_\xi\cdot \nabla_x$, assumptions A1 and A2 imply
\be\non
\big|\p_{t^k}F(\xi,t)\big| \leq C\lan x\ran^{-M_F-k} \lan x_\xi \ran^{k(M_\om - 1)} \leq C_1 \lan x_\xi \ran^{-M_F-k + k(M_\om - 1)},
\ee 
where in the last inequality we have used Lemma~\ref{l:prem}(2).
Then, in view of \eqref{mu_prop}, for $0\leq k \leq 2$ 
\be\lbl{Phi_est}
\big|\p_{t^k}\Phi(\xi,t)\big| \leq \max_{0\leq i \leq k} C\lan x_\xi \ran^{-M_F - i + i(M_\om-1) + (k-i)(M_\om-2)}
= C\lan x_\xi \ran^{-M_F + k(M_\om-2)}. 
\ee
Denote $\eta_{\xi,\theta}(t):=\theta^2g_\xi^2(t) + \eps_\xi^2$.
According to \eqref{g_xi}, for $0\leq k \leq 2$
\be\lbl{eta_est}
\Big|\frac{d^k}{dt^k} \eta_{\xi,\theta}(t)\Big| = \theta^2\Big|\frac{d^k}{dt^k} g^2_{\xi,\theta}(t)\Big|
\leq C\theta^2 \lan x_\xi \ran^{k(M_\om - 2)}. 
\ee
Combining estimates \eqref{Phi_est}, \eqref{eta_est} and the estimate $|\eta_{\xi,\theta}(t)|\geq C\theta^2$ following from \eqref{g_xi}, we get
\be\non
\Big|\frac{\p_{t^2}\Phi} {\eta_{\xi,\theta}}\Big|, \qu
 \Big|\frac{\p_{t}\Phi \, \eta_{\xi,\theta}'} {\eta^2_{\xi,\theta}}\Big|, \qu
  \Big|\frac{\Phi \, \eta_{\xi,\theta}''} {\eta^2_{\xi,\theta}}\Big|, \qu
   \Big|\frac{\Phi \, \big(\eta_{\xi,\theta}'\big)^2} {\eta^3_{\xi,\theta}}\Big| 
 	\le
C \theta^{-2}  \lan x_\xi\ran^{-M_F + 2(M_\om-2)}.
 \ee
Then, 
$
\big| f_{\xi,\theta}''(t) \big| \leq
C\theta^{-2}\lan x_\xi\ran^{-M_F + 2(M_\om-2)},
$
so, by \eqref{2309}, 
\be\lbl{J-froz}
\big|J_\nu(\xi) - J_\nu^0(\xi)\big| \leq 
C\theta_\xi \lan x_\xi\ran^{-M_F + 2(M_\om-2)} \leq C \lan x_\xi\ran^{-M_F + M_\om-2}.
\ee
Let $I_\nu^{f}$ denotes the integral \eqref{fin_form} in which the internal integral $J_\nu(\xi)$ is replaced by $J_\nu^0(\xi)$,
\be\lbl{I_nu^f}
I_\nu^{f} =  \int_{\Sigma_0\sm B_\de} \frac{J_\nu^0(\xi)\,m(d\xi)}{|N_{
		\xi}|^3}. 
\ee
According to \eqref{J-froz}, 
\be\non
\big| \blan I_\nu, V_\de \bran - I_\nu^f \big| 
\leq C \int_{\Sigma_0\sm B_\de} \frac{\lan x_\xi\ran^{-M_F + M_\om-2}\, m(d\xi)}{|N_{
		\xi}|^3} \leq C_1\chi_{d,4}(\de),
\ee
due to Corollary~\ref{c:integration} and assumption A4. Then, representation \eqref{I_nu_sum} and estimates \eqref{first_int}, \eqref{second_int}, \eqref{third_int_appr} imply that to prove the theorem it suffices to establish the desired asymptotic \eqref{main} in which the integral $I_\nu$ is replaced by $I_\nu^f$.

\subsection{Study of the integral $I_\nu^f$}

Recall that the integral $I_\nu^f$ is defined in 
 \eqref{I_nu^f}, where the function $J_\nu^0(\xi)$ is given by~\eqref{J_nu^0}. 
 Applying the inequality
$
\ds{0<\frac\pi2 - \arctan\frac1\ga < \ga}$ 
with $\ga = \eps_\xi/\theta_\xi$, which holds for $0<\ga\leq 1/2$,  we get
\be\lbl{arctg_approx}
\Big|\frac{\pi F(\xi,0)}{\eps_\xi} - J^0_\nu(\xi)\Big| <\frac{2 F(\xi,0)}{\theta_\xi} 
\ee
if 
\be\lbl{eps/theta}
\frac{\eps_\xi}{\theta_\xi}  = \frac{\nu\lan x_\xi\ran^{M_\om-2}}{|N_\xi|^{2}\Theta} \leq \frac12.
\ee
Assume first $|x_\xi|\ge 1$. Then, according to assumption A3,
$
\eps_\xi / \theta_\xi \leq
C \Theta^{-1} \nu |x_\xi|^{M_\om-2 - 2m_\om},
$
so for  sufficiently small $\nu$ the inequality \eqref{eps/theta} is satisfied if  $M_{cr}
:= M_\om-2m_\om-2 \leq 0$  or $M_{cr}>0$ and
\be\lbl{area_>}
|x_\xi|\leq C_{cr}(\Theta\nu^{-1})^{1/M_{cr}}
\ee
with appropriate constant $C_{cr}>0$.

In the case $|x_\xi|<1$ with $\xi\in\Sigma_0\sm B_\de$, according to \eqref{N_xi-lower}, 
$$
\frac{\eps_\xi}{\theta_\xi} \leq  C\frac{ \nu}{\Theta |x_\xi|^{2}} \leq C \Theta^{-1} \nu \de^{-2}  = C\Theta^{-1} \al^{-2},
$$
where we recall the definition \eqref{de-al} of $\de$.
Choosing $\al = \sqrt{2C}\Theta^{-1/2}$, we see that \eqref{eps/theta} is satisfied.

Let us introduce a subset $\Sigma_0^<\ni\xi$  of  $\Sigma_0$ such that in $\Sigma_0^<\sm B_\de$  the inequality \eqref{eps/theta} is fulfilled. 
More specifically, 
if $M_{cr}>0$ we define (cf. \eqref{area_>}) 
$$
\Sigma_0^< = \big\{\xi\in\Sigma_0:\; |x_\xi|\leq C_{cr}(\Theta\nu^{-1})^{1/M_{cr}}\big\} 
$$
and if $M_{cr}\leq 0$, we set $\Sigma_0^< = \Sigma_0$. 
We denote $\Sigma_0^> = \Sigma_0 \sm \Sigma_0^<$ and write
\be\non
I_\nu^f =  
\Big(\int_{\Sigma_0^<\sm B_\de} + \int_{\Sigma_0^>}\Big)  \frac{ m(d\xi)}{|N_{\xi}|^3}J_\nu^0(\xi),
\ee
where we assume $\nu$ to be so small that $\Sigma_0^> \cap B_1 = \emptyset.$
 By \eqref{J_nu^0}, the integral over $\Sigma_0^>$ is bounded by
\be\lbl{864658}
C\nu^{-1}\int_{\Sigma_0^>} \frac{m(d\xi)}{|N_\xi|}\, |F(\xi,0)|
\leq
C (C_{cr}\Theta)^{-1}\int_{\Sigma_0^>}  \frac{m(d\xi)}{|N_\xi|}\, |x_\xi|^{M_{cr}}|F(\xi,0)|. 
\ee
According to Corollary~\ref{c:integration} and assumptions A1 and A4, the latter integral is bounded by a $\nu$-independent constant. 

 It remains to study the integral over $\Sigma_0^<\sm B_\de$. Denote by $K_\nu$ the integral obtained from the latter by approximating $J^0_\nu(\xi)$ via \eqref{arctg_approx}, i.e.
\be\non
K_\nu= \pi\int_{\Sigma_0^<\sm B_\de}\frac{F(\xi, 0)\,m(d\xi)}{\eps_\xi|N_{\xi}|^3} = \pi\nu^{-1}\int_{\Sigma_0^<\sm B_\de}\frac{F(\xi, 0)\,m(d\xi)}{|N_{\xi}|}.
\ee
According to Corollary~\ref{c:integration} this integral converges,
as well as the integral \eqref{asymp}
$$
I_0 = \pi  \int_{\Sigma_0}\frac{F(\xi, 0)\,m(d\xi)}{|N_{\xi}|} =  \pi\int_{\Sigma_0} \frac{F(x)}{\big|\nabla \om(x)\big|}\, d_\Sigma x. 
$$        
Since for $\xi\in \Sigma_0^<$ the inequality \eqref{arctg_approx} is satisfied, 
\be\lbl{ap_151}
\Big|\int_{\Sigma_0^<\sm B_\de} \frac{m(d\xi)}{|N_{\xi}|^3}J_\nu^0(\xi)
 - K_\nu\Big| \leq 
 2\int_{\Sigma_0^<\sm B_\de}\frac{|F(\xi, 0)|\,m(d\xi)}{\theta_\xi|N_{\xi}|^3} \leq C_1\chi_{d,4} (\de),
\ee
again by Corollary~\ref{c:integration}. 
To conclude the proof of the theorem it remains to note that, by~\eqref{N_xi-lower}, 
\be\non
|K_\nu- \nu^{-1} I_0| \leq 
\pi\nu^{-1}\Big( \int_{\Sigma_0\cap B_\de} + \int_{\Sigma_0^>}\Big)\frac{|F(\xi,0)|\,m(d\xi)}{|N_\xi|}.
\ee
The first integral in the r.h.s. above is bounded by 
$ C\nu^{-1} \int_{\Sigma_0\cap B_\de} |x_\xi|^{-1}\,m(d\xi)\leq  C_1\nu^{-1} \de^{2} \leq C_2$
according to Lemma~\ref{l:int_zero}(1) 
since $d\geq 4$,
while the second integral coincides with the l.h.s. of \eqref{864658}, so is bounded by a constant.
Proof of Theorem~\ref{t:main} is finished.

\bigskip

{\bf Acknowledgement.} I am deeply grateful to Sergei B. Kuksin for our discussions.

\noindent This work was supported by the Russian Science Foundation under grant no. 19-71-30012.

\end{document}